# LiMn$_{1-x}$Fe$_x$PO$_4$ Nanorods Grown on Graphene Sheets for Ultra-High Rate Performance Lithium Ion Batteries

Hailiang Wang,[1,§] Yuan Yang,[2,§] Yongye Liang,[1] Li-Feng Cui,[2] Hernan Sanchez Casalongue,[1] Yanguang Li,[1] Guosong Hong,[1] Yi Cui[*,2] and Hongjie Dai[*1].

[1]Department of Chemistry and Laboratory for Advanced Materials, and [2]Department of Materials Science and Engineering, Stanford University, Stanford, CA 94305, USA.

[§] These two authors contributed equally to this work

* Correspondence to hdai@stanford.edu and yicui@stanford.edu

Following the successful utilization of LiFePO$_4$ as a novel cathode material for rechargeable lithium batteries, interest and efforts have grown in the research of another olivine structured material LiMnPO$_4$ due to its higher operating potential voltage and energy density. However, high rate performance for LiMnPO$_4$-based cathode materials has been challenging due to its extremely low electrical and ionic conductivities. Here, we develop a synthesis of Fe-doped LiMnPO$_4$ (LiMn$_{0.75}$Fe$_{0.25}$PO$_4$) nanorods directly bonded on graphene sheets (reduced from graphene oxide) to render LiMn$_{0.75}$Fe$_{0.25}$PO$_4$ nanorods superior electrical conductivity. The LiMn$_{0.75}$Fe$_{0.25}$PO$_4$ nanorod morphology is unique to materials grown on graphene over those grown in free solution, and is ideal for fast Li$^+$ diffusion with the diffusion path of [010] crystallographic axis along the short radial direction (~20-30nm) of the nanorods. These together lead to ultrafast discharge within ~30-40 seconds without using a high carbon content.

**With high columbic efficiency above 99.5% at a high operating voltage, our LiMn$_{0.75}$Fe$_{0.25}$PO$_4$ nanorods/graphene hybrid exhibits the best rate performance among all doped LiMnPO$_4$ cathode materials for Lithium ion batteries.**

Olivine-type lithium transition-metal phosphates LiMPO$_4$ (M = Fe, Mn, Co, or Ni) have been intensively investigated as promising cathode materials for rechargeable lithium ion batteries (LIBs) due to their high capacity, excellent cycle life, thermal stability, environmental benignity and low cost.[1-18] However, the inherently low ionic and electrical conductivities of LiMPO$_4$ seriously limit Li$^+$ insertion and extraction and charge transport rates in these materials. In recent years, these obstacles have been overcome for LiFePO$_4$ by reducing the size of LiFePO$_4$ particles to nanoscale and applying conductive surface coatings such as carbon, which leads to commercially viable LiFePO$_4$ cathode materials.[5-12]

Compared to LiFePO$_4$, LiMnPO$_4$ is an attractive cathode material due to its higher Li$^+$ intercalation potential of 4.1V vs. Li$^+$/Li (3.4V for LiFePO$_4$), providing ~20% higher energy density than LiFePO$_4$ for LIBs.[13-18] Importantly, the 4.1V intercalation potential of LiMnPO$_4$ is compatible with most of the currently used liquid electrolytes.[13-18] However, the electrical conductivity of LiMnPO$_4$ is lower than the already insulating LiFePO$_4$ by 5 orders of magnitude,[13-18] making it challenging to achieve high capacity at high rates for LiMnPO$_4$ using methodologies developed for LiFePO$_4$.[13-18] Doping LiMnPO$_4$ with Fe has been pursued to enhance conductivity and stability of the material in its charged form.[19-24] Recently, Martha et al. have obtained improved capacity and rate performance for carbon coated LiMn$_{0.8}$Fe$_{0.2}$PO$_4$ nanoparticles synthesized by a high temperature solid state reaction.[22]

Graphene is an ideal substrate for growing and anchoring insulating materials for energy storage applications due to its high conductivity, light weight, high mechanical strength and structural flexibility.[25-27] The electrochemical performance of various electrode materials can be significantly boosted by rendering them conducting with graphene sheets.[28-31] Recent works have shown improved specific capacity and rate capability of simple oxide nanomaterials ($Mn_3O_4$, $Co_3O_4$, and $Fe_3O_4$) grown on graphene as LIB anode materials.[29-31] However, it remains wide open to grow nanocrystals on graphene sheets in solution for materials with more sophisticated compositions and structures such as $LiMn_{1-x}Fe_xPO_4$, which is a promising but extremely insulating cathode material for LIBs.

Here we devise a two-step approach for synthesis of $LiMn_{1-x}Fe_xPO_4$ nanorods on reduced graphene oxide sheets stably suspended in solution. Fe-doped $Mn_3O_4$ nanoparticles were first selectively grown onto graphene oxide by controlled hydrolysis. The oxide nanoparticle precursors then reacted solvothermally with Li and phosphate ions and were transformed into $LiMn_{1-x}Fe_xPO_4$ on the surface of reduced graphene oxide sheets. With a total content of 26 wt% conductive carbon, the resulting hybrid of nanorods and graphene showed high specific capacity and unprecedentedly high power rate for $LiMn_{1-x}Fe_xPO_4$ type of cathode materials. Stable capacities of 132mAh/g and 107mAh/g were obtained at high discharge rates of 20C and 50C, 85% and 70% of the capacity at C/2 (155mAh/g) respectively. This affords LIBs with both high energy and high power densities. This is also the first synthesis of $LiMn_{0.75}Fe_{0.25}PO_4$ nanorods, with an ideal crystal shape and morphology for fast $Li^+$ diffusion along the radial [010] direction of the nanorods.

Figure 1 shows our two step solution phase reaction scheme for synthesizing $LiMn_{0.75}Fe_{0.25}PO_4$ nanorods on reduced graphene oxide. The first step was to selectively grow oxide nanoparticles at 80°C on mildly oxidized graphene oxide (mGO) stably suspended in a solution. Controlling the hydrolysis rate of $Mn(Ac)_2$ and $Fe(NO_3)_3$ by adjusting the $H_2O$/$N,N$-dimethylformamide (DMF) solvent ratio and the reaction temperature afforded selective and uniform coating of ~10nm nanoparticles of Fe-doped $Mn_3O_4$ on the GO sheets without free growth of nanoparticles in solution. Importantly, our mGO was made by a modified Hummers method, with which a 6 times lower concentration of $KMnO_4$ oxidizer was used to afford milder oxidation of graphite.[32-35] The resulting mGO sheets contained lower oxygen content than Hummers' GO [~15% vs. ~30% measured by X-ray photoelectron spectroscopy (XPS) and Auger spectroscopy] and showed higher electrical conductivity when chemically reduced than Hummer's GO. Our mGO still contained sufficient functional groups such as carboxyl, hydroxyl and epoxide groups for nucleating and anchoring oxide nanoparticles on the surface.[29]

The second step reaction transformed the Fe-doped $Mn_3O_4$ nanoparticles on mGO into $LiMn_{0.75}Fe_{0.25}PO_4$ nanorods (Figure 1), by reacting with LiOH and $H_3PO_4$ solvothermally at 180°C. Ascorbic acid ($V_C$) was added to reduce Fe(III) to Fe(II), and reduce mGO as well.[36,37] The reduction is highly effective due to the solvothermal condition (Oxygen content was reduced to ~4% as revealed by XPS).[35] This afforded highly conducting reduced graphene oxide sheets (rmGO) with the formation of $LiMn_{0.75}Fe_{0.25}PO_4$ nanorods atop. The electrical conductivity measured from pellets of $LiMn_{0.75}Fe_{0.25}PO_4$/rmGO hybrid was measured to be ~0.1-1 S/cm, $10^{13}$-$10^{14}$ times higher than pure $LiMnPO_4$.

Scanning electron microscopy (SEM) and transmission electron microscopy (TEM) revealed selective growth of single-crystalline $LiMn_{0.75}Fe_{0.25}PO_4$ nanorods on rmGO (Figure 2a, 2c), with rod length of ~50-100nm and diameter of ~20-30nm. High resolution TEM showed the crystal lattice fringes throughout the entire $LiMn_{0.75}Fe_{0.25}PO_4$ nanorod formed on rmGO (Figure 2d), indicating the $LiMn_{0.75}Fe_{0.25}PO_4$ nanorods were single crystals. The X-ray diffraction peaks of the nanorods were slightly shifted to larger 2θ angles compared to pure $LiMnPO_4$ due to Fe doping (Figure 2b) [19-24] homogeneously in a solid solution (Figure 2d). Energy dispersive spectroscopy (EDS) showed that the Mn/Fe ratio was ~3 in the nanorods. The amount of rmGO in the hybrid was found to be ~20% by mass through thermal analysis.

After the first step reaction, we observed uniform coating of Mn/Fe oxide on mGO, a result of metal cation adsorption on the functional groups on mGO and hydrolysis:

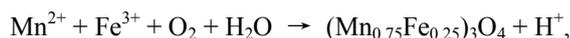

$$Mn^{2+} + Fe^{3+} + O_2 + H_2O \rightarrow (Mn_{0.75}Fe_{0.25})_3O_4 + H^+,$$

which afforded small oxide nanoparticles on mGO. The reaction was carried out in an open system exposed to air. The second step solvothermal reaction between the Mn/Fe oxide nanoparticles with Li and phosphate ions afforded $LiMn_{0.75}Fe_{0.25}PO_4$ specifically on the surface of mGO accompanied by mGO reduction:

$(Mn_{0.75}Fe_{0.25})_3O_4 + Li^+ + PO_4^{3-} + H^+ + V_C(reduced) \rightarrow LiMn_{0.75}Fe_{0.25}PO_4 + V_C(oxidized) + H_2O$.

By systematically varying the reaction time, we observed gradual transformation of precursor particles into phosphate nanorods on rmGO.

In a control experiment, the same synthesis steps without mGO added produced irregularly-shaped $LiMn_{0.75}Fe_{0.25}PO_4$ particles without the desired nanorod morphology. Thus, our results suggested that mildly oxidized graphene sheets present a unique substrate for growing nanocrystals into well defined morphologies such as nanoplates[26] and nanorods. While the functional groups on mGO allow for adsorption of cations and nanoparticle nucleation to achieve uniform precursor coating in the first step of reaction, the conjugated graphitic regions of rmGO (formed by reduction of mGO in solvothermal condition) interact with surface species weakly to promote the formation of well defined shapes of nanocrystals in the second step of reaction. The resulting nanorods could be bonded to rmGO via Mn/Fe-O-C bonds at the oxygen remaining sites and via van der Waals interactions with the aromatic regions of rmGO.

The nanoscale sizes (length ~50-100nm, width ~20-30nm) of the $LiMn_{0.75}Fe_{0.25}PO_4$ nanorods were desirable due to decreased transport length for Li ions and electrons.[8,14] High resolution TEM analysis revealed that the long axes of the nanorods grown on rmGO were along the [001] axis of the $LiMn_{0.75}Fe_{0.25}PO_4$ crystal structure [Figure 3c, 3d, inter-planar distance of (001) was ~0.47nm]. More importantly, the [010] axis, which is the $Li^+$ diffusion channel direction in olivine type of crystals (Figure 3a),[8,13,16] was along one of the short axes of the nanorods [Figure 3c, 3d, inter-planar distance of (100) was ~1.04nm]. This could be ideal for fast $Li^+$ insertion and extraction with the $LiMn_{0.75}Fe_{0.25}PO_4$ nanorods/rmGO hybrid as the cathode of a lithium ion battery. Small diameters of the $LiMn_{0.75}Fe_{0.25}PO_4$ nanorods result in large surface areas and thus facilitate fast Li ion transport at the interface between the $LiMn_{0.75}Fe_{0.25}PO_4$ nanorods and the electrolyte. The $Li^+$ channels along the radial direction of the nanorods also favor rapid Li ion diffusion within the nanorods.

Coin cells were made to test the electrochemical performance of our $LiMn_{0.75}Fe_{0.25}PO_4$ nanorods/rmGO hybrid material (after annealing at 600℃ for 1h) as the cathode (at a loading of ~3mg/cm$^2$) and with a Li foil as the anode. Figure 4a showed the typical charge and discharge curves of our material at a rate of C/2 (C rate based on the theoretical specific capacity of $LiMn_{0.75}Fe_{0.25}PO_4$, where a 1C rate corresponded to a current density of 170mA/g). The charge curve showed two voltage plateaus at ~3.6V and ~4.2V, corresponding to the oxidation from Fe(II) to Fe(III) and from Mn(II) to Mn(III), respectively.[19-24] The capacity of the 3.6V plateau was ~1/3 of that of the 4.2V plateau, consistent with the Fe/Mn ratio of the $LiMn_{0.75}Fe_{0.25}PO_4$ nanorods. The discharge curve also showed two voltage plateaus at ~4.0V and ~3.5V due to reduction from Mn(III) to Mn(II) and from Fe(III) to Fe(II), respectively.

Figure 4b showed the discharge curves at various rates for our $LiMn_{0.75}Fe_{0.25}PO_4$ nanorods/rmGO cathode material. Charging was done at a C/2 constant current to 4.25V followed by a constant voltage charging at 4.25V with a cut-off current of C/20.

At a C/2 discharge rate and 2.0V cutoff voltage, our material showed a specific capacity of 155mAh/g based on the mass of the active material $LiMn_{0.75}Fe_{0.25}PO_4$ (Figure 4c), which is 91% of the theoretical capacity. The capacity remained high at 153mAh/g at a discharge rate of 2C. At a high discharge rate of 20C, the capacity was still as high as 132mAh/g. For even higher discharge rates, we obtained a capacity of 107mAh/g at 50C, and 65mAh/g at 100C (corresponding to a discharge time of 36 seconds). A cut-off voltage of 2.0V was used for all the discharge rates above (Figure 4b).

We obtained the best rate performance using similar measurement and voltage cut-off conditions as prior work on pure and doped $LiMnPO_4$ cathode materials (see Figure 4e for comparison, we used a 2.7V cut-off voltage in this case). Notably, the rate performance of our $LiMn_{0.75}Fe_{0.25}PO_4$ nanorods/rmGO hybrid was also comparable to some of the best reported cathode materials (including $LiFePO_4$, $LiCoO_2$ and $LiMn_2O_4$) with ultrahigh rate capabilities.[10,12,38-40] Moreover, the content of conductive carbon in our work was only 26% including both rmGO and carbon black, which was much lower than used previously for discharging rates $\geq$ 100C (45% carbon or higher).[10,38,40] Further, our material exhibited good cycle life, showing little capacity decay (~1.9%) every one hundred cycles from the 11th to the 100th cycle (Figure 4d). The Coulombic efficiency (CE) was typically higher than 99.0%. After 50 cycles, CE was above 99.5%, suggesting little side chemical reaction or stable solid electrolyte interphase (SEI) formation between electrolyte and $LiMn_{0.75}Fe_{0.25}PO_4$ nanorods/rmGO hybrid.[13]

The outstanding electrochemical performance of our $LiMn_{0.75}Fe_{0.25}PO_4$/rmGO hybrid material for Li battery was attributed to the intimate interactions between the nanorods and the underlying reduced graphene oxide sheets, and the small size and ideal nanorod morphology and crystallographic orientation of the $LiMn_{0.75}Fe_{0.25}PO_4$ nanocrystals. In a control experiment, a physical mixture of $LiMn_{0.75}Fe_{0.25}PO_4$ nanoparticles and rmGO sheets showed much lower specific capacities even at low rates (75 mAh/g at C/10 and 44 mAh/g at 1C rate). Capacity of the mixture further decreased to less than 2mAh/g at a discharge rate of 50C, clearly showing that the capacity contribution of rmGO at high rate is negligible. Electrochemical impedance spectroscopy (EIS) provided strong evidence of significantly improved charge transport in the cathode made of $LiMn_{0.75}Fe_{0.25}PO_4$ nanorods directly grown on rmGO over the simple physical mixture. Cyclic voltammetry (CV) measurement also showed much better kinetics of $LiMn_{0.75}Fe_{0.25}PO_4$ nanorods directly grown on rmGO compared to the physical mixture of $LiMn_{0.75}Fe_{0.25}PO_4$ and rmGO.

In summary, we developed the first synthesis of complex single crystalline nanomaterials on highly conducting mildly oxidized graphene sheets with desired size and morphology. The resulting high electrical conductivity and low ionic resistance led to excellent rate performance for the otherwise extremely insulating $LiMn_{0.75}Fe_{0.25}PO_4$ cathode material. This work opens the door to complex hybrid materials design and engineering with graphene for advanced energy storage.

**Acknowledgement**

This work was supported by Office of Naval Research, NSF CHE-0639053 and the LDRD Funding from the DOE SLAC Accelerator National Laboratory. H.W and Y.Y acknowledge financial support from Stanford Graduate Fellowship.

**References**

1. Whittingham, M. S. Lithium batteries and cathode materials. *Chem. Rev.* **104**, 4271-4301 (2004).

2. Ellis, B. L., Lee, K. T. & Nazar, L. F. Positive electrode materials for Li-ion and Li-batteries. *Chem. Mater.* **22**, 691-714 (2010).

3. Bruce, P. G., Scrosati, B. & Tarascon, J. Nanomaterials for rechargeable lithium batteries. *Angew. Chem. Int. Ed.* **47**, 2930-2946 (2008).

4. Murugan, A. V., Muraliganth, T., Ferreira, P. J. & Manthiram, A. Dimensionally modulated, single-crystalline $LiMPO_4$ (M= Mn, Fe, Co, and Ni) with nano-thumblike shapes for high-power energy storage. *Inorg. Chem.* **48**, 946-952 (2009).

5. Padhi, A. K., Nanjundaswamy, K. S. & Goodenough, J. B. Phospho-olivines as positive-electrode materials for rechargeable lithium batteries. *J. Electrochem. Soc.* **144**, 1188–1194 (1997).

6. Yamada, A., Chung, S. C. & Hinokuma, K. Optimized $LiFePO_4$ for lithium battery cathodes. *J. Electrochem. Soc.* **148**, A224-A229 (2001).

7. Huang, H., Yin, S. & Nazar, L. F. Approaching theoretical capacity of $LiFePO_4$ at room temperature at high rates. *Electrochem. Solid State Lett.* **4**, A170-A172 (2001).

8. Delacourt, C., Poizot, P., Levasseur, S. & Masquelier, C. Size effects on carbon-free $LiFePO_4$ powders. *Electrochem. Solid State Lett.* **9**, A352-A355 (2006).

9. Wang, Y., Wang, Y., Hosono, E., Wang, K. & Zhou, H. The Design of a LiFePO$_4$/carbon nanocomposite with a core–shell structure and its synthesis by an in situ polymerization restriction method. *Angew. Chem. Int. Ed.* **47**, 7461-7465 (2008).

10. Kang, B. & Cedar, G. Battery materials for ultrafast charging and discharging. *Nature*, **458**, 190-193 (2009).

11. Oh, S. W., Myung, S., Oh, S., Oh, K. H., Amine, K., Scrosati, B. & Sun, Y. Double carbon coating of LiFePO$_4$ as high rate electrode for rechargeable lithium batteries. *Adv. Mater.* 2010, DOI: 10.1002/adma.200904027.

12. Zhang, W. Comparison of the rate capacities of LiFePO4 cathode materials. *J. Electrochem. Soc.* **157**, A1040-A1046 (2010).

13. Wang, L., Zhou, F. & Ceder, G. Ab Initio Study of the Surface Properties and Nanoscale Effects of LiMnPO$_4$. *Electrochem. Solid State Lett.* **11**, A94-A96 (2008).

14. Drezen, T., Kwon, N., Bowen, P., Teerlinck, I., Isono, M. & Exnar, I. Effect of particle size on LiMnPO$_4$ cathodes. *J. Power Sources* **174**, 949-953 (2007).

15. Martha, S. K., Markovsky, B., Grinblat, J., Gofer, Y., Haik, O., Zinigrad, E., Aurbach, D., Drezen, T., Wang, D., Deghenghi, G. & Exnar, I. LiMnPO$_4$ as an advanced cathode material for rechargeable lithium batteries. *J. Electrochem. Soc.* **156**, A541-A552 (2009).

16. Choi, D., Wang, D., Bae, I., Xiao, J., Nie, Z., Wang, W., Viswanathan, V. V., Lee, Y. J., Zhang, J., Graff, G. L., Yang, Z. & Liu, J. LiMnPO$_4$ nanoplate grown via solid-state reaction in molten hydrocarbon for Li-ion battery cathode. *Nano Lett.* **10**, 2799-2805 (2010).

17. Bakenov, Z. & Taniguchi, I. Physical and electrochemical properties of LiMnPO$_4$/C composite cathode prepared with different conductive carbon. *J. Power Sources* **195**, 7445-7451 (2010).

18. Kang, B. & Cedar, G. Electrochemical performance of LiMnPO$_4$ synthesized with off-stoichiometry. *J. Electrochem. Soc.* **157**, A808-A811 (2010).

19. Chang, X., Wang, Z., Li, X., Zhang, L., Guo, H. & Peng, W. Synthesis and performance of LiMn$_{0.7}$Fe$_{0.3}$PO$_4$ cathode material for lithium ion batteries. *Mater. Res. Bull.* **40**, 1513-1520 (2005).

20. Mi, C. H., Zhang, X. G., Zhao, X. B. & Li, H. L. Synthesis and performance of LiMn$_{0.6}$Fe$_{0.4}$PO$_4$ nano-carbon webs composite cathode. *Mater. Sci. Eng. B* **129**, 8-13 (2006).

21. Zaghib, K., Mauger, A., Gendron, F., Massot, M. & Julien, C. M. Insertion properties of LiFe$_{0.5}$Mn$_{0.5}$PO$_4$ electrode materials for Li-ion batteries. *Ionics* **14**, 371-376 (2008).

22. Martha, S. K., Grinblat, J., Haik, O., Zinigrad, E., Drezen, T., Miners, J. H., Exnar, I., Kay, A., Markovsky, B. & Aurbach, D. LiMn$_{0.8}$Fe$_{0.2}$PO$_4$: an advanced cathode material for rechargeable lithium batteries. *Angew. Chem. Int. Ed.* **48**, 8559-8563 (2009).

23. Zhang, B., Wang, X., Liu, Z., Li, H. & Huang, X. Enhanced electrochemical performances of carbon coated mesoporous LiFe$_{0.2}$Mn$_{0.8}$PO$_4$. *J. Electrochem. Soc.*

**157**, A285-A288 (2010).

24. Wang, Y., Zhang, D., Yu, X., Cai, R., Shao, Z., Liao, X. & Ma, Z. Mechanoactivation-assisted synthesis and electrochemical characterization of manganese lightly doped LiFePO$_4$. *J. Alloys Compd.* **492**, 675-680 (2010).

25. Park, S. & Ruoff, R. S. Chemical methods for the production of graphenes. *Nature Nanotechnol.* **4**, 217-224 (2009).

26. Wang, H., Robinson, J. T., Diankov, G. & Dai, H. Nanocrystal growth on graphene with various degrees of oxidation. *J. Am. Chem. Soc.* **132**, 3270-3271 (2010).

27. Liang, Y., Wang, H., Casalongue, H. S., Chen, Z. & Dai, H. TiO$_2$ nanocrystals grown on graphene as advanced photocatalytic hybrid materials. *Nano Res.* **3**, 701-705 (2010).

28. Wang, H., Casalongue, H. S., Liang, Y. & Dai, H. Ni(OH)$_2$ nanoplates grown on graphene as advanced electrochemical pseudocapacitor materials. *J. Am. Chem. Soc.* **132**, 7472-7477 (2010).

29. Wang, H., Cui, L., Yang, Y., Casalongue, H. S., Robinson, J. T., Liang, Y., Cui, Y. & Dai, H. Mn$_3$O$_4$−graphene hybrid as a high-capacity anode material for lithium ion batteries. *J. Am. Chem. Soc.* **132**, 13978-13980 (2010).

30. Yang, S., Cui, G., Pang, S., Cao, Q., Kolb, U., Feng, X., Maier, J. & Mullen, K. Fabrication of cobalt and cobalt oxide/rmGO composites: towards high-performance anode materials for lithium ion batteries. *ChemSusChem* **3**, 236 − 239 (2010).

31. Zhang, M., Lei, D., Yin, X., Chen, L., Li, Q., Wang, Y. & Wang, T. Magnetite/rmGO composites: microwave irradiation synthesis and enhanced cycling and rate performances for lithium ion batteries. *J. Mater. Chem.* **20**, 5538-5543 (2010).

32. Hummers, W. S. & Offeman, R. E. Preparation of graphitic oxide. *J. Am. Chem. Soc.* **80**, 1339-1339 (1958).

33. Wang, H., Wang, X., Li, X. & Dai, H. Chemical self-assembly of graphene sheets. *Nano Res.* **2**, 336– 342 (2009).

34. Sun, X., Liu, Z., Welsher, K., Robinson, J. T., Goodwin, A., Zaric, S. & Dai, H. Nano-graphene oxide for cellular imaging and drug delivery. *Nano Res.* **1**, 203– 212 (2008)

35. Wang, H., Robinson, J., Li, X. & Dai, H. Solvothermal reduction of chemically exfoliated graphene sheets. *J. Am. Chem. Soc.* **131**, 9910-9911 (2009).

36. Fernandez-Merino, M. J., Guardia, L., Paredes, J. I., Villar-Rodil, S., Solis-Fernandez, P., Martinez-Alonso, A. & Tascon, J. M. D. Vitamin C is an ideal substitute for hydrazine in the reduction of graphene oxide suspensions. *J. Phys. Chem. C* **114**, 6426-6432 (2010).

37. Gao, J., Liu, F., Liu, Y., Ma, N., Wang, Z. & Zhang, X. Environment-friendly method to produce graphene that employs vitamin C and amino acid. *Chem. Mater.* **22**, 2213-2218 (2010).

38. Okubo, M., Hosono, E., Kim, J., Enomoto, M., Kojima, N., Kudo, T., Zhou, H. & Honma, I. Nanosize effect on high-rate Li-ion intercalation in $LiCoO_2$ electrode. *J.*

*Am. Chem. Soc.* **129**, 7444-7452 (2007).

39. Shaju, K. M. & Bruce, P. G. A stoichiometric nano-LiMn$_2$O$_4$ spinel electrode exhibiting high power and stable cycling. *Chem. Mater.* **20**, 5557-5562 (2008).

40. Hosono, E., Kudo, T., Honma, I., Matsuda, H. & Zhou, H. Synthesis of single crystalline spinel LiMn$_2$O$_4$ nanowires for a lithium ion battery with high power density. *Nano Lett.* **9**, 1045-1051 (2009).

**Figures**

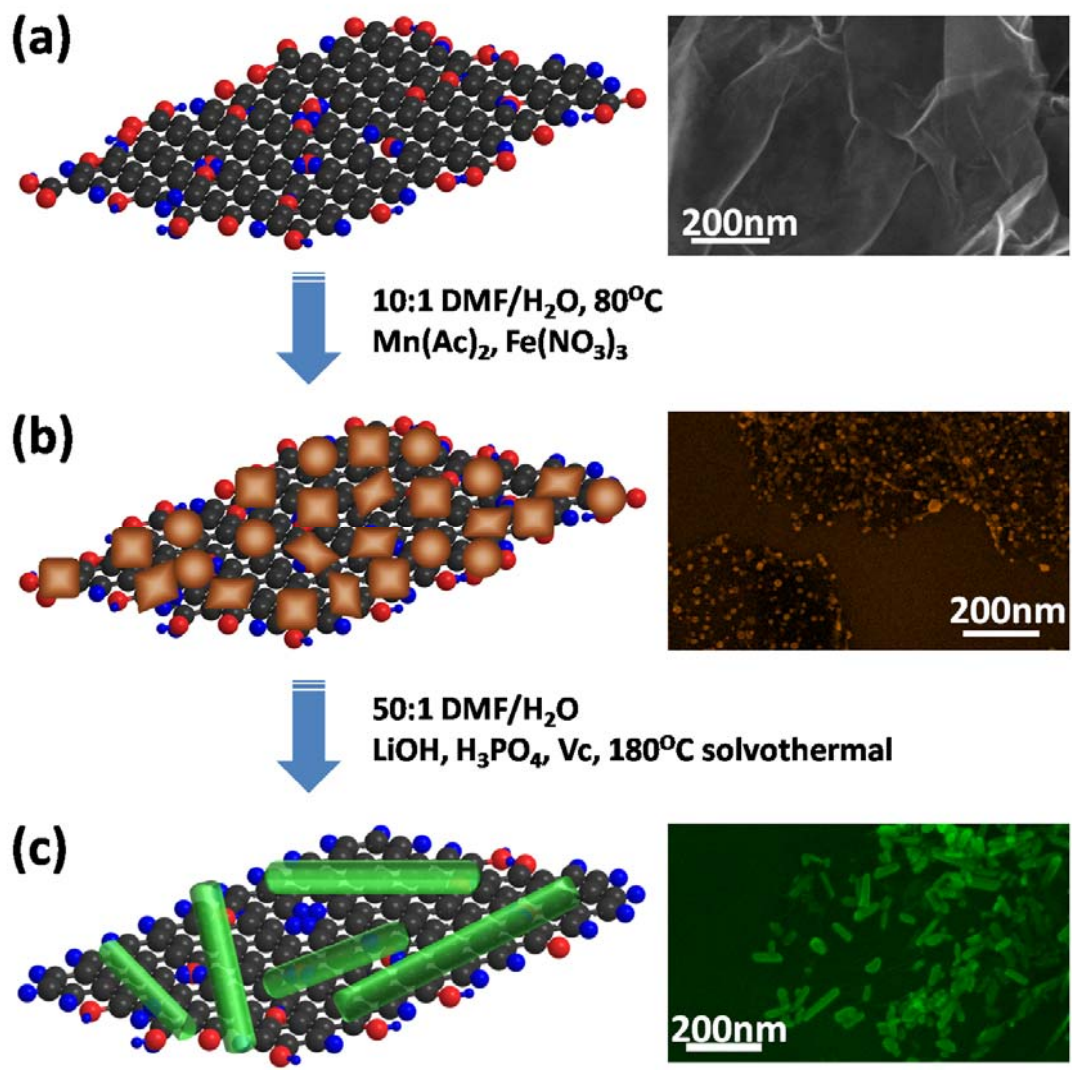

**Figure 1. Schematic LiMn$_{0.75}$Fe$_{0.25}$PO$_4$ nanorod growth on rmGO.** (a) Ball-stick (dark gray balls: carbon atoms; blue balls: hydrogen atoms; red balls: oxygen atoms) schematic structure (left) and SEM image (right) of mGO. (b) Schematic structure (left) and SEM image (right) of Fe-doped Mn$_3$O$_4$ precursor nanoparticles grown on mGO after the first step of reaction at 80℃. (c) Schematic structure (left) and SEM image (right) of LiMn$_{0.75}$Fe$_{0.25}$PO$_4$ nanorods grown on rmGO after the second step of solvothermal reaction at 180℃.

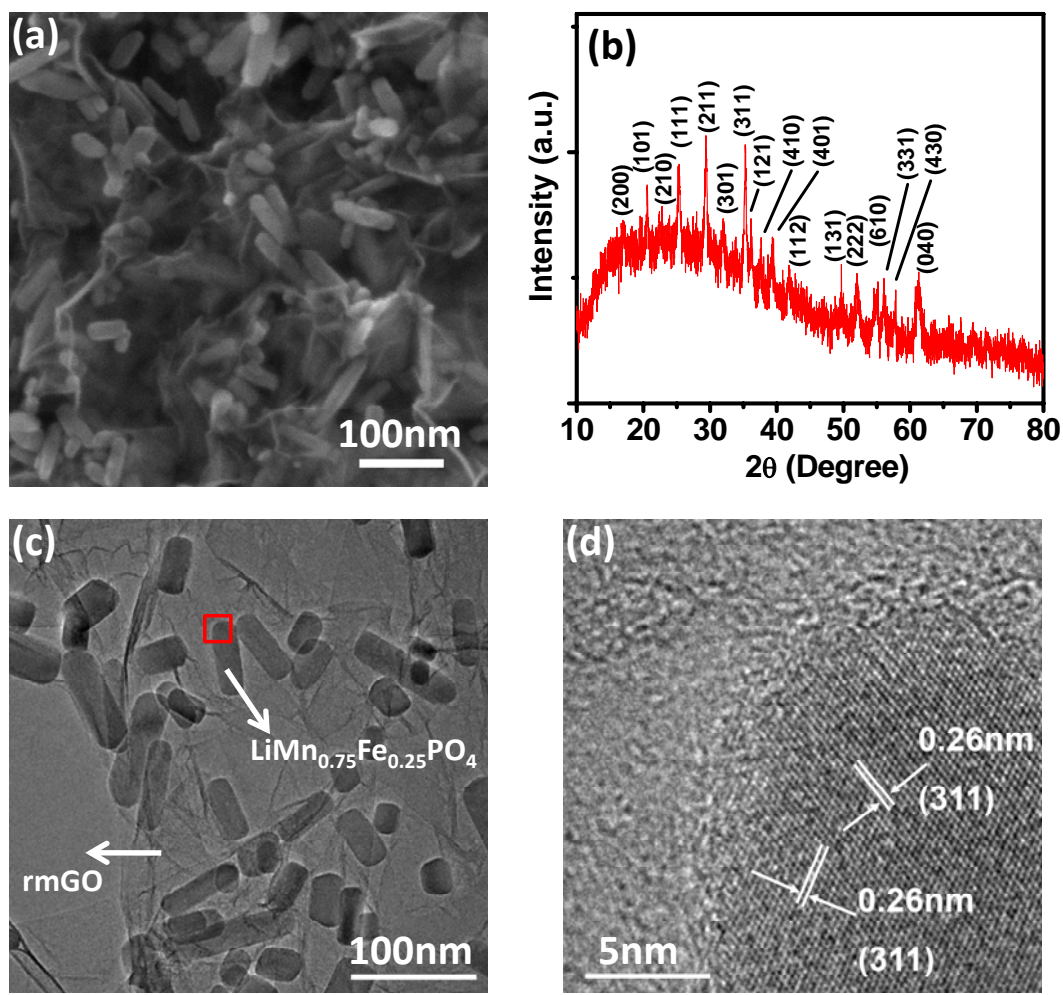

**Figure 2. LiMn$_{0.75}$Fe$_{0.25}$PO$_4$ nanorods grown on rmGO.** (a) SEM image of LiMn$_{0.75}$Fe$_{0.25}$PO$_4$/rmGO hybrid. (b) XRD spectrum of a packed thick film of LiMn$_{0.75}$Fe$_{0.25}$PO$_4$/rmGO. (c) TEM image of LiMn$_{0.75}$Fe$_{0.25}$PO$_4$/rmGO. (d) High resolution TEM image of an individual LiMn$_{0.75}$Fe$_{0.25}$PO$_4$ nanorod on rmGO (red lined area in c).

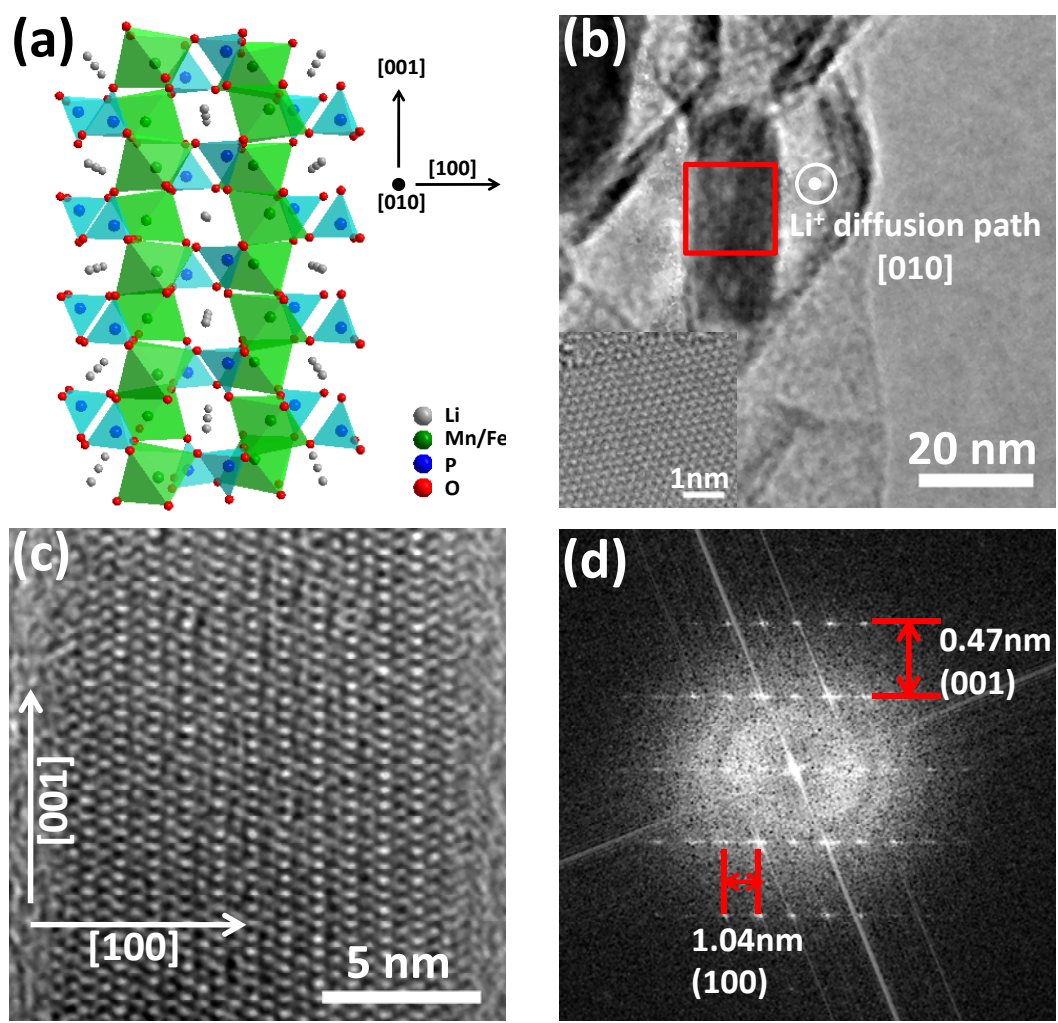

**Figure 3. High resolution TEM analysis of an individual LiMn$_{0.75}$Fe$_{0.25}$PO$_4$ nanorod grown on rmGO.** (a) Schematic crystal structure of LiMn$_{0.75}$Fe$_{0.25}$PO$_4$. The Li diffusion channels are clearly shown to be along the [010] direction (perpendicular to the figure plane). (b) TEM image of a LiMn$_{0.75}$Fe$_{0.25}$PO$_4$ nanorod grown on rmGO. Inset shows an aberration corrected TEM lattice image of an rmGO sheet. (c) High resolution TEM image of the LiMn$_{0.75}$Fe$_{0.25}$PO$_4$ nanorod on rmGO (red lined area in b). (d) Fast Fourier transform of the lattice structure in c. It was clearly revealed that the growth direction of the LiMn$_{0.75}$Fe$_{0.25}$PO$_4$ nanorod was [001], and the Li diffusion channel [010] was one of the short axes of the nanorods (perpendicular to the figure plane).

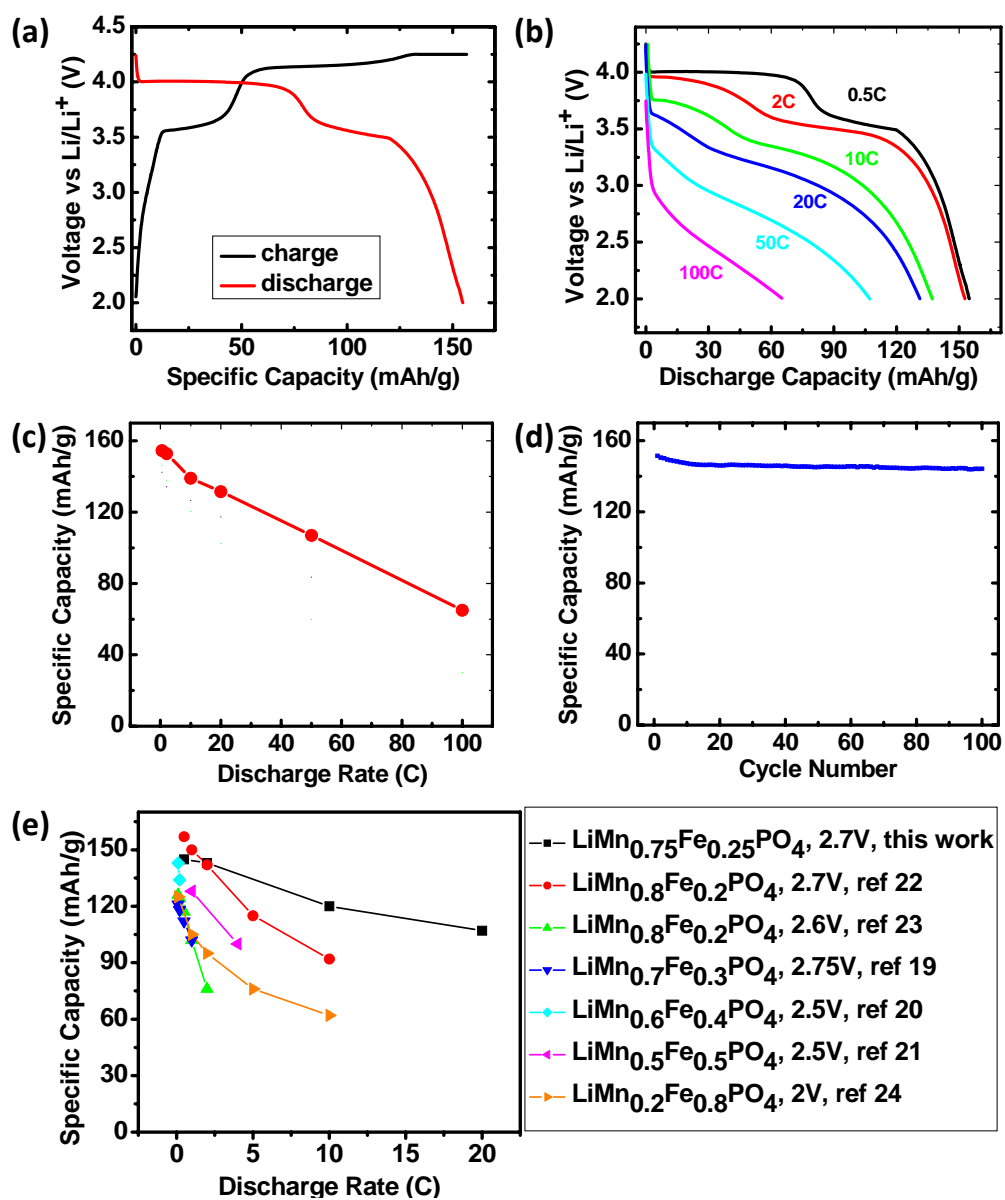

**Figure 4. Electrochemical characterizations of the LiMn$_{0.75}$Fe$_{0.25}$PO$_4$ nanorod cathode grown on rmGO.** (a) Representative charge (black) and discharge (red) curves of LiMn$_{0.75}$Fe$_{0.25}$PO$_4$ nanorods grown on rmGO at the rate of C/2. (b) Discharge curves of LiMn$_{0.75}$Fe$_{0.25}$PO$_4$ nanorods on rmGO at various C rates. (c) Specific discharge capacities of LiMn$_{0.75}$Fe$_{0.25}$PO$_4$ nanorods on rmGO at various C rates. The discharge cut-off voltage was 2.0V vs Li$^+$/Li. (d) Capacity retention of LiMn$_{0.75}$Fe$_{0.25}$PO$_4$ nanorods on rmGO at the rate of C/2. (e) Comparison of rate capability of LiMn$_{0.75}$Fe$_{0.25}$PO$_4$ nanorods grown on rmGO with other Fe-doped LiMnPO$_4$ cathode materials. The discharge cut-off voltages are listed in the legend on the right.